\documentstyle[11pt,moriond,epsfig]{article}

\def\Journal#1#2#3#4{{#1} {\bf #2}, #3 (#4)}

\def\PLB{{\em Phys. Lett.}  B}
\def\PRL{ {\em Phys. Rev. Lett.}}
\def\PRD{{\em Phys. Rev.} D}

\def\be{\begin{equation}}
\def\ee{\end{equation}}
\def\bea{\begin{eqnarray}}
\def\eea{\end{eqnarray}}

\def\babar{BABAR}
\def\jpsi{J/\psi}
\def\psitwos{\psi(2S)}
\def\chic{\chi_{c1}}
\def\U4S{\Upsilon(4S)}

\def\pstar{$p^*$}
\def\ifb{\mathrm{fb^{-1}}}

\def\Bogg{B^0 \to \gamma \gamma}

\newcommand{\etal}{{\em et al.}}

\begin{document}
\begin{flushright}
SLAC-PUB-9008 \\
BABAR-PROC-01/28
\end{flushright}

\vspace*{3cm}
\title{STUDY OF CHARMONIUM PRODUCTION AND ELECTROWEAK PENGUINS WITH BABAR}

\author{ Vuko Brigljevi\'c }

\address{Lawrence Livermore National Laboratory, 7000 East Ave,\\
Livermore CA 94550, USA \\ \vspace*{0.2cm}
{representing the BABAR Collaboration}}

\maketitle\abstracts{
We report measurements of charmonium resonances ($\jpsi$, $\psitwos$, $\chic$) using 
about 25 $\ifb$ of data collected by the BABAR detector around the $\U4S$ resonance. 
We present measurements of inclusive charmonium production of charmonium in $B$ decays
and from the continuum, as well as exclusive branching ratios of $B$ mesons into charmonium
final states. We present also a measurement of the $B^0 \to K^{0*} \gamma$ branching ratio and 
a search for the decay $\Bogg$.
}

\section{Introduction}

We present measurements of charmonium production and processes involving electroweak
penguins in $e^+ e^-$ collisions at the $\U4S$ resonance, using data taken
by the \babar\ detector~\cite{babar} at the PEP-II $B$ factory in 1999 and 2000, which consist
of  $20.7\, \ifb$ accumulated at the $\Upsilon(4S)$ resonance, and  $2.6\, \ifb$ taken off-resonance 
at an energy 0.04 GeV below the peak. This sample corresponds to $22.7 \cdot 10^{6}$ 
$\U4S \to B\bar{B}$ decays.

\section{Charmonium production}

We reconstruct decays of the charmonium resonances $\jpsi$, $\psitwos$ \
and $\chic$. We reconstruct $\jpsi$\ and $\psitwos$\ through their decay into 
two electrons or two muons; $\psitwos$\ is also reconstructed through
the decay $\jpsi \pi^+ \pi^-$, while $\chic$\ is reconstructed through the decay 
into $\jpsi \, \gamma$.  As examples, the signals for the decays $\jpsi \to e^+e^-$,
$\chic \to \jpsi \, \gamma\ (\jpsi\to \mu^+\mu^-)$ and
 $\psitwos \to \jpsi \pi^+ \pi^- (\jpsi \to e^+e^-)$ are shown in 
Fig.~\ref{fig:charmonium-signals}. 

\subsection{Inclusive Charmonium studies}\label{subsec:prod}

Charmonium mesons may be produced: a) as a product of a $B$ meson 
decay; b) as a direct product of the decay of $\U4S$; c) in the fragmentation process 
of a continuum $e^+e^- \to q \bar{q}$ event (prompt production); d) through Inital State Radiation
(ISR).

We isolate charmonium mesons from B decays by looking at $B\bar{B}$-like events
and by requiring the charmonium momentum in the center of mass frame, $p^*$,
to be below the kinematic limit for $B$ decays,  less than 2 GeV/c 
for $\jpsi$ and less than 1.6 GeV/c for $\psitwos$. Results for inclusive branching ratios
of $B$ mesons into charmonium mesons are listed in Table 1.

\begin{table}[tb]
\begin{center}
\begin{tabular}{|l|c|c|} \hline
Meson              &   $\mu \mu / ee$    & $\mathcal{B}(B \to $ Meson X) [\%] \\  \hline
$\jpsi$            &  $0.995 \pm 0.036$  & $ 1.044 \pm 0.013 \pm 0.028 $  \\ 
$\jpsi$ direct     &  $0.999 \pm 0.045$  & $ 0.789 \pm 0.010 \pm 0.034 $  \\
$\psitwos$         &  $0.93  \pm 0.15$   & $ 0.275 \pm 0.020 \pm 0.029 $  \\
$\chi_{c1}$        &  $1.09  \pm 0.21$   & $ 0.378 \pm 0.034 \pm 0.026 $  \\
$\chi_{c1}$ direct &  $1.11  \pm 0.23$   & $ 0.353 \pm 0.034 \pm 0.024 $  \\
$\chi_{c2}$        &  $0.78  \pm 0.68$   & $ 0.137 \pm 0.058 \pm 0.012 $  \\
$\chi_{c2}$ limit  &                     & $ < 0.21 $ at 90\% C.L.  \\  \hline
\end{tabular}
\caption{Inclusive Branching Ratios of $B$ mesons into charmonium mesons. }
\end{center}
\label{tab:inclBF}
\end{table}

\begin{figure}[tb]
\begin{center}
\psfig{figure=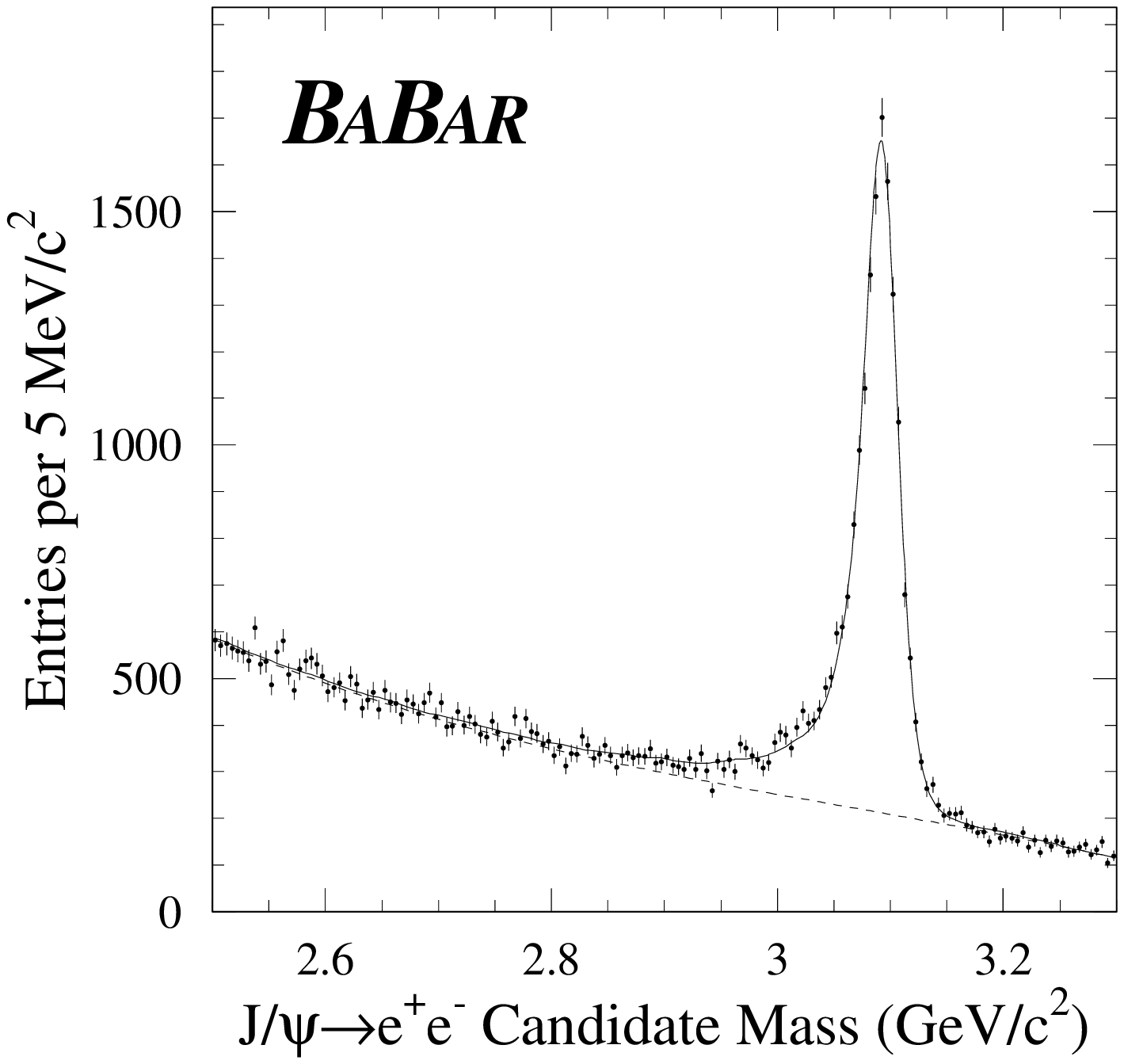,width=0.3\textwidth}
\psfig{figure=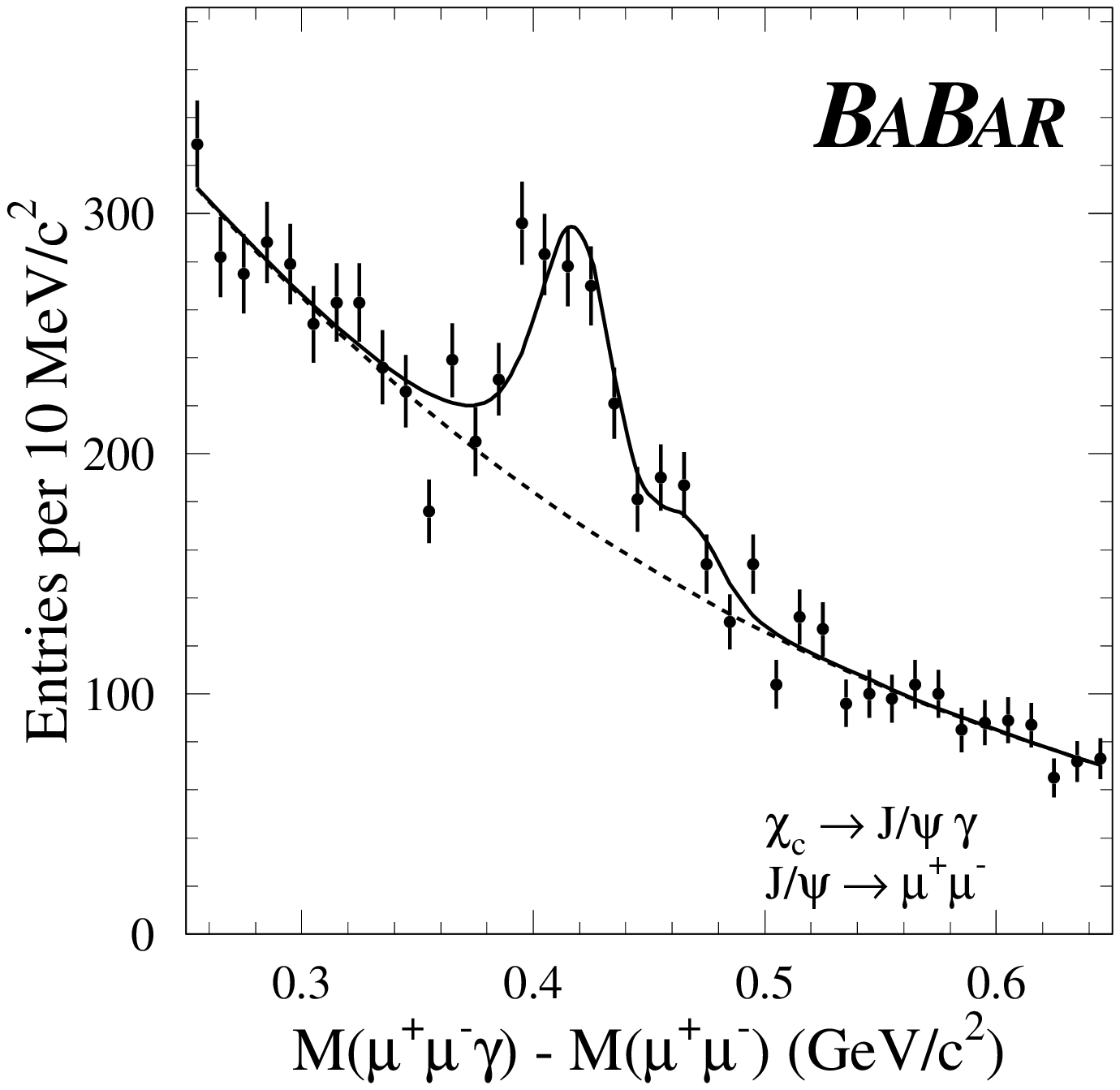 ,width=0.3\textwidth}
\psfig{figure=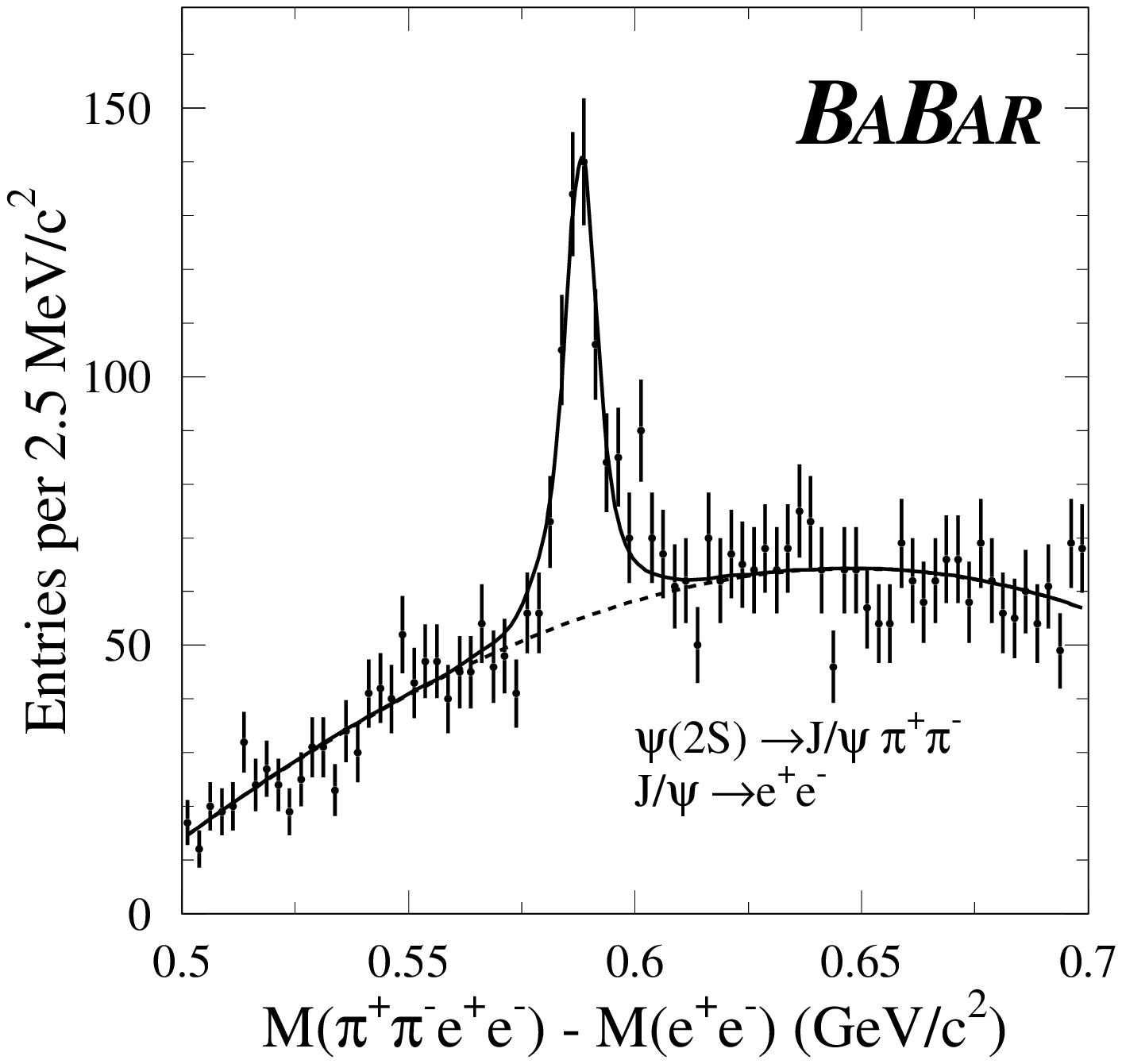 ,width=0.3\textwidth}
\end{center}
\caption{Charmonium signals, from left to right: $M(e^+ e^-)$ for  $\jpsi \ to  e^+ e^-$ candidates,
         $M(\mu \mu \gamma) - M(\mu \mu)$ for $\chic \to \jpsi(\mu\mu) \gamma$ candidates and
	 $M(e^+ e^- \pi^+ \pi^-)  - M (e^+ e^-)$ for   $ \psitwos \to \jpsi(e^+ e^-) \pi^+ \pi^-$
         candidates.
\label{fig:charmonium-signals}}
\end{figure}

We mesure the $\jpsi$ polarization by fitting the helicity distribution. The helicity
angle, $\Theta_H$, is the angle, measured in the $\jpsi$ rest frame, between the positively
charged lepton and the $\jpsi$ flight direction in the $B$ meson rest frame \footnote{The $B$ meson 
rest frame is approximated by the $\U4S$ rest frame.}. The distribution of $u=\cos \Theta_H$
can be written in terms of a polarization parameter 
$\alpha$: $h(u) = 3 (1 + \alpha u^2) / [2(\alpha+3)]$, where $\alpha=0$ indicates the distribution
is unpolarized, $\alpha=1$ is transversely polarized and $\alpha=-1$ is longitudinally polarized.
We find $\alpha = -0.424 \pm 0.023$ for $\jpsi$ mesons from B decays.

The $\jpsi$ production in the continuum is of  particular interest due to the possible contribution 
of $c\bar{c}$ pairs created in a color octet state, which would 
enhance prompt $\jpsi$ production~\cite{octetTheory-Yuan,octetTheory-Schuler}.
To eliminate background from $ B \to \jpsi X$ in the on-peak data sample, we require 
the $\jpsi$ momentum in the $\U4S$ rest frame to be greater than 2 GeV/c.
To suppress ISR production of $\jpsi$ and $\psitwos$ and 
two photon production of $\chi_{c2}$, we require at least 3 quality tracks  
with $0.41<\theta<2.54$,  the visible energy of the event be greater than 5 GeV and the ratio
of the second to the zeroth Fox-Wolfram moment, $R_2$, to be smaller than 0.5. We then
study the production and decay properties of these prompt $\jpsi$ mesons. The distribution 
of the signal in $\cos \Theta^*$ has been extracted and fit with $1+A \cos^2 \Theta^*$.
Color octet and color singlet models have very different predictions for the value of $A$: 
at high $p^*$ values, color octet models predict $0.6 < A < 1.0$ while the color 
singlet model predicts $A \approx -0.8$~\cite{octet-anglePrediction}. We measure 
$A = 1.5 \pm 0.6$ for \pstar$>3.5$ GeV/c, clearly favoring the presence of color octet 
contributions. We also measure the polarization for prompt $\jpsi$ to be $\alpha = -0.73 \pm 0.09$.


\subsection{Exclusive Charmonium decays}

We look for candidate $B$ mesons by combining the reconstructed charmonium mesons with light
meson candidates. Two kinematic variables are used to isolate the $B$ meson signal: the difference
$\Delta E$ between the reconstructed energy of the candidate and the beam energy in the 
$\U4S$ rest frame , and the beam energy substituted mass $m_{ES}$, defined as 
$m_{ES} = \sqrt{E^{*\ 2}_{beam} - p^{*\ 2}_{B}}$, where $p^{*}_{B}$ is the momentum
of the reconstructed $B$ candidate in the $\U4S$ rest frame. We determine
branching fractions for 14 exclusive $B$ meson  charmonium decay modes\cite{PRD-ExclCharm}, listed 
in Table 2. We report, in particular, the first observation of the decay 
$B^0 \to \chi_{c1} K^{*0}\, (\to K^+ \pi^-)$. 

\begin{figure}[t]
\begin{center}
\begin{minipage}{9.6cm}
\hspace{0.3cm}
\label{tab:tagging}
\begin{center}
\small
\begin{tabular}{|ll|c|} \hline
Channel                    &                          & BF / $10^{-4}$ \\ \hline
$ B^0 \to \jpsi K^0$       &  $K^0_S \to \pi^+ \pi^-$ &  $ 8.5 \pm 0.5 \pm 0.6$   \\
                           &  $K^0_S \to \pi^0 \pi^0$ &  $ 9.6 \pm 1.5 \pm 0.7$   \\
                           &  $K^0_L$                 &  $ 6.8 \pm 0.8 \pm 0.8$   \\
                           &  All                     &  $ 8.3 \pm 0.4 \pm 0.5$   \\
$ B^+ \to \jpsi K^+$       &                          &  $10.1 \pm 0.3 \pm 0.5$   \\
$ B^0 \to \jpsi \pi^0$     &                          &  $0.20 \pm 0.06 \pm 0.02$   \\
$ B^0 \to \jpsi K^{*0}$    &                          &  $12.4 \pm 0.5 \pm 0.9$   \\
$ B^+ \to \jpsi K^{*+}$    &                          &  $13.7 \pm 0.9 \pm 1.1$   \\
$ B^0 \to \psitwos K^0$    &                          &  $ 6.9 \pm 1.1 \pm 1.1$   \\
$ B^+ \to \psitwos K^+$    &                          &  $ 6.4 \pm 0.5 \pm 0.8$   \\

$ B^0 \to \chi_{c1} K^0$      &                          &  $ 5.4 \pm 1.4 \pm 1.1$   \\
$ B^+ \to \chi_{c1} K^+$      &                          &  $ 7.5 \pm 0.8 \pm 0.8$   \\
$ B^0 \to \chi_{c1} K^{*0}$   &                          &  $ 4.8 \pm 1.4 \pm 0.9$   \\

\hline
\end{tabular}
\end{center}
\begin{minipage}{9.6cm}
\vspace*{0.3cm}
\footnotesize
Table 2: Measured branching fractions for exclusive decays of $B$ mesons involving
charmonium. The first error is statistical and the second systematic.
\end{minipage}
\end{minipage}
\hspace{0.1cm}
\begin{minipage}{6cm}
\begin{center}
\psfig{figure=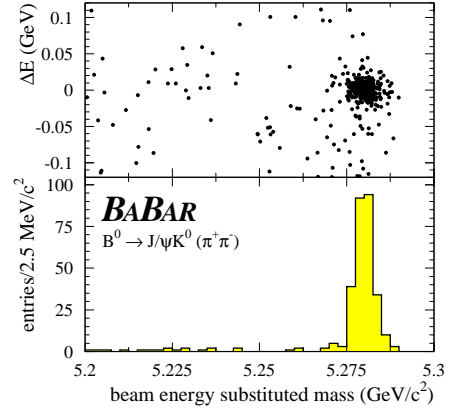,width=1.0\textwidth}
\vspace*{-0.66cm}
\caption{\label{sin2b-asym} Energy difference $\Delta E$ vs energy substituted mass 
$m_{ES}$ for the golden $CP$ channel $B^0 \to \jpsi K^0_S(\pi^+\pi^-)$.}
\end{center}
\end{minipage}
\end{center}
\end{figure}

\section{Electroweak Penguins}

Electroweak penguins could be particularly sensitive to the presence of new physics and they 
could be low energy probes of new phenomena at a much higher energy scale. 
We present in the following the measurement of the decay $B \to K^* \gamma$
and a search for the decay $B^0 \to \gamma \gamma$.

\subsection{ Decay $ B \to K^* \gamma$}

The decay $B \to K^* \gamma$ proceeds by the electroweak penguin transition $ b \to s \gamma$.
We reconstruct this decay in the mode $K^* \to K^+ \pi^-$. The radiative photon 
candidate is found by looking for a cluster in the electromagnetic calorimeter consistent 
with a photon shower  and with an energy between 1.5 and 4.5 Gev in the laboratory, and 2.30 
and and 2.85 Gev in the center of mass frame. The $K^+$ and $\pi^-$ candidates are 
identified thanks to the DIRC, an internally-reflecting ring-imaging Cherenkov detector 
(DIRC), requiring that the cone of light must be consistent with the pion or kaon hypothesis, 
which leads in a correct $K/\pi$ assignment in 97\% of the cases.

The main background is from continuum $q\bar{q}$ production with the high-energy photon
originating from initial state radiation or from $\pi^0$ or $\eta$ decays. We exploit
event topology differences between signal and background to reduce the continuum
contribution. The first variable used to achieve that is $|\cos \Theta^*_T|$, where
$\Theta^*_T$ is the angle, measured in the center of mass frame,  between the photon 
candidate and the thrust vector of the event excluding the $B$ daughter candidates. While
the distribution of $|\cos \Theta^*_T|$ is flat between 0 and 1 for the signal, it is
peaked at 1 for the continuum background. Thus, we require $|\cos \Theta^*_T|<0.8$. We further
suppress backgrounds using the angle of the $B$ candidate's direction with respect to the
beam axis, $\Theta_B^*$, and the helicity angle of the $K^*$ decay, $\Theta_H^*$, defined
as the angle between the $K^\pm$ momentum vector computed in the rest frame of the 
$K^*$ and the $K^*$ momentum vector in the parent $B$ menson rest frame. This distribution 
follows a $\sin^2\Theta^*_H$ distribution for signal and is approximately flat for $q\bar{q}$ 
background. The same is true for the $B$ candidate direction with respect to the beam axis. 
We require $|\cos \Theta^*_B|<0.80$ and $|\cos \Theta^*_H|<0.75$. The signal 
is shown in Fig.~\ref{fig:penguins} We find a yield of $139 \pm 13$ events and we derive the 
branching fraction $\mathcal{B}$$(B^0 \to K^* \gamma) = (4.39 \pm 0.41 \pm 0.27)\cdot 10^{-5}$.

This sample is used to search for $CP$ violating charge asymmmetries by 
constructing $A_{CP} = [ (\bar{B} \to \bar{K}^* \gamma) - (B \to K^* \gamma) ]
 / [ (\bar{B} \to \bar{K}^* \gamma) +  (B \to K^* \gamma) ]$. The flavour of the underlying
$b$ quark is tagged by the charge of the $K^\pm$ in the decay. We constrain 
$A_{CP} = -0.035 \pm 0.094 \pm 0.022$.

\begin{figure}
\begin{tabular}{ccc}
\parbox[t]{0.3\textwidth} {
\epsfig{figure=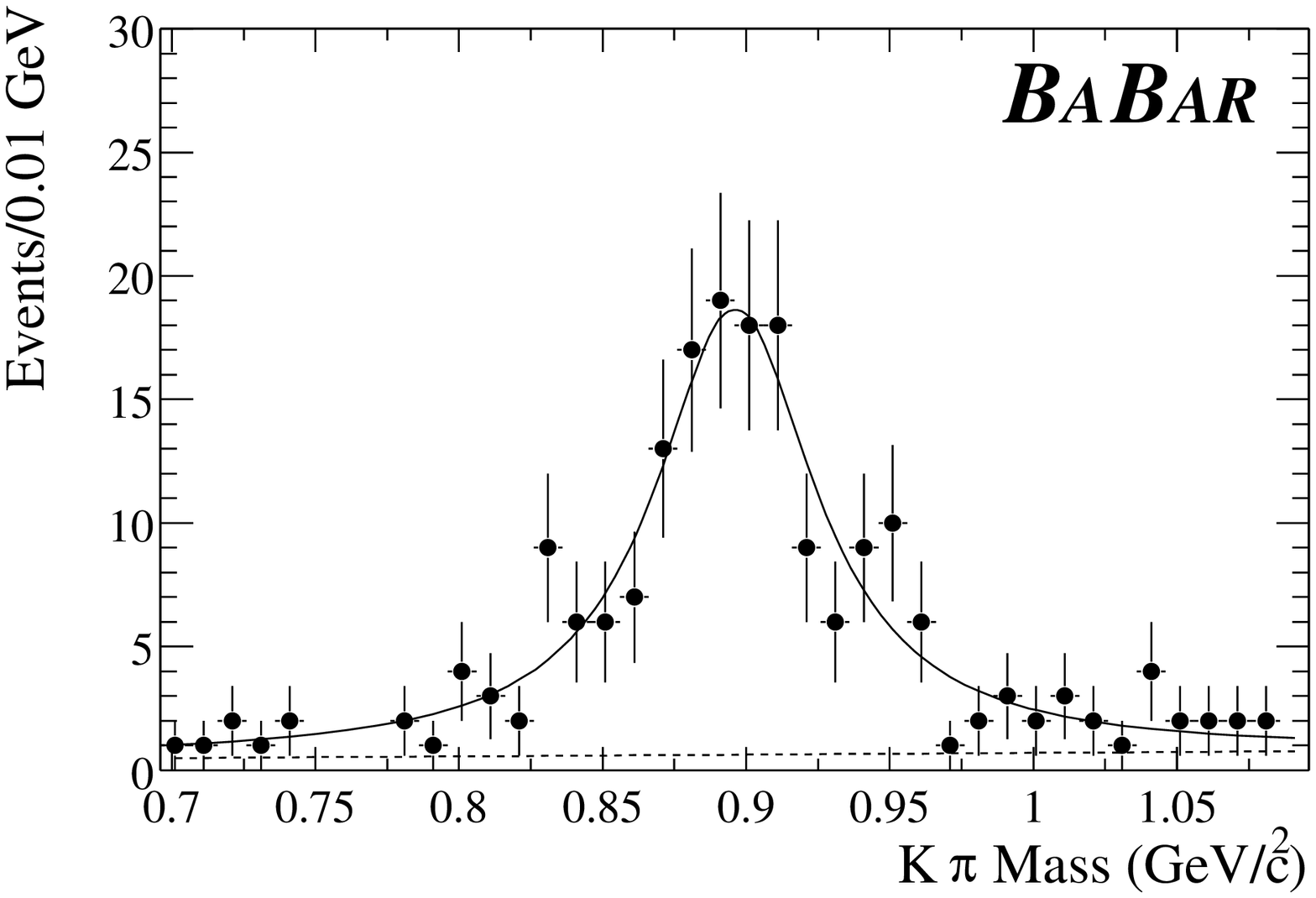,width=0.3\textwidth}
} &
\parbox[t]{0.3\textwidth} {
\epsfig{figure=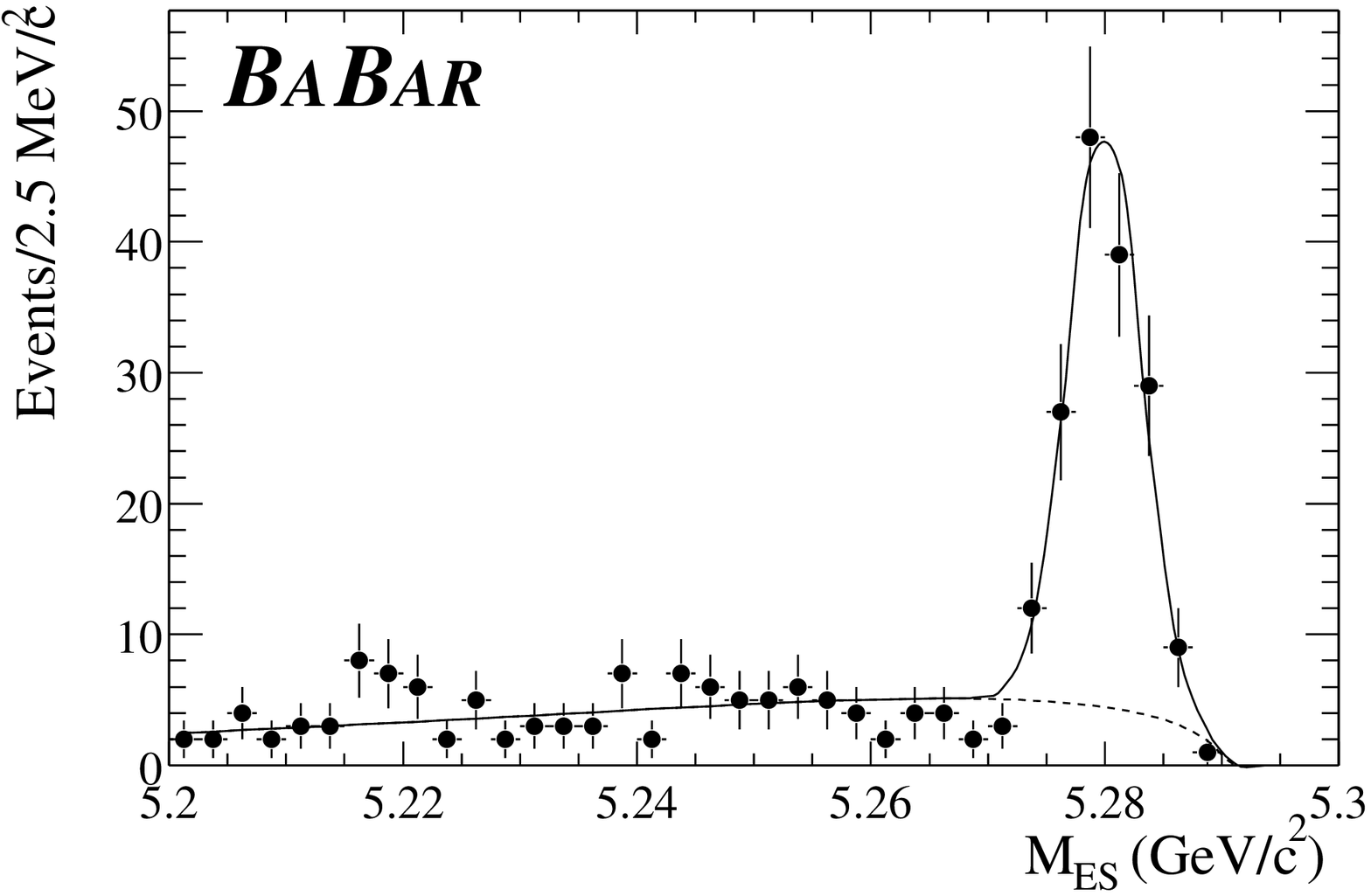,width=0.3\textwidth}
} &
\parbox[t]{0.3\textwidth} {
\epsfig{figure=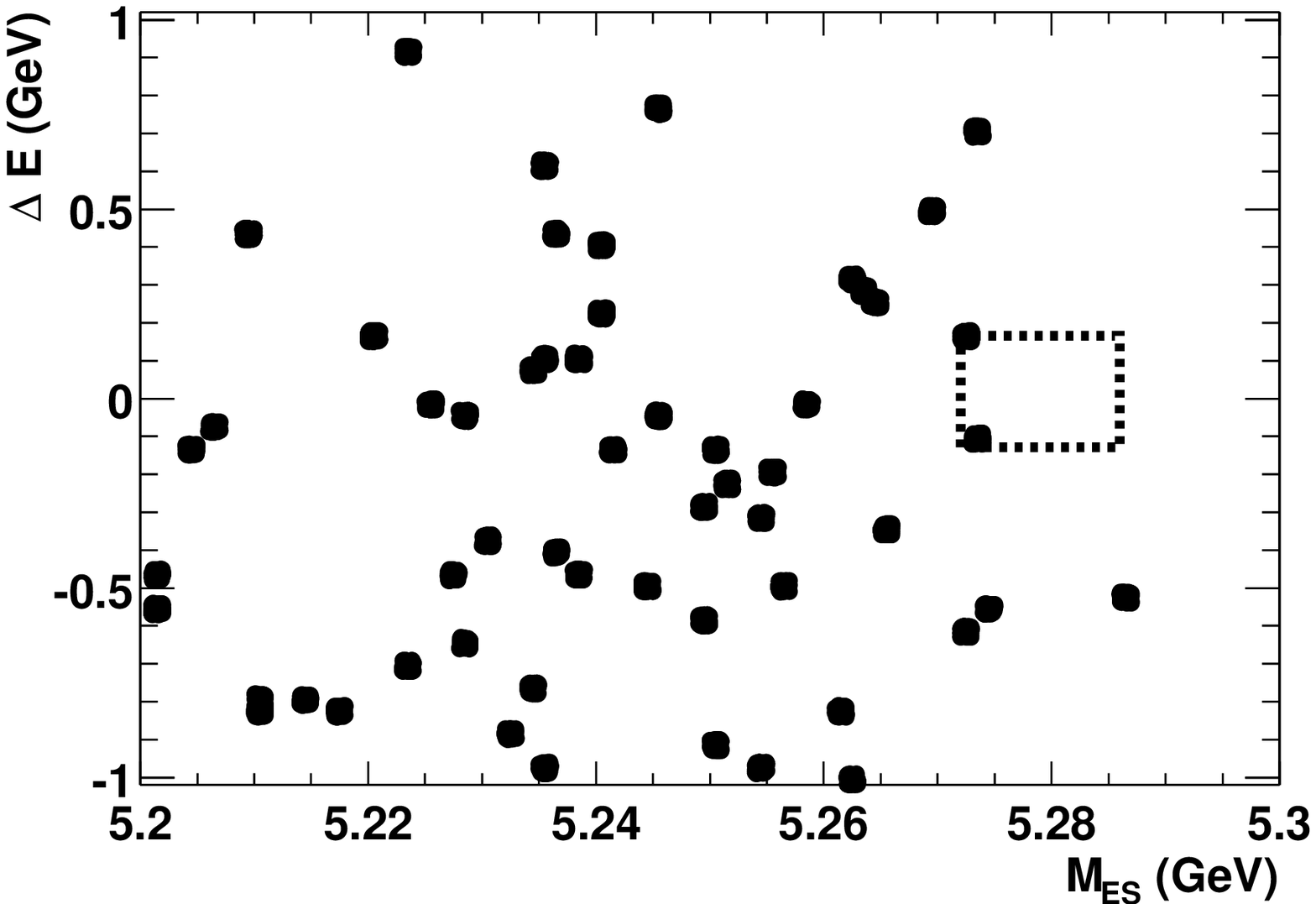,width=0.3\textwidth}
}
\end{tabular}

\caption{The fist two plots, from left to right, show the $K\pi$ mass and the energy susbstituted
         mass for $B \to K^* \gamma$ candidates; the third plot shows $\Delta E$ and $m_{ES}$ for
         the $\Bogg$ candidates.}
\label{fig:penguins}

\end{figure}

\subsection{ Decay $\Bogg$}

In the Standard Model, the decay $B^0 \to \gamma \gamma$ proceeds via a second order 
weak transition including gluonic penguings, followed by annihilation. Standard Model 
predictions for the branching fraction of these effective flavor-changing neutral 
current processes range from 0.1 to $2.3 \cdot 10^{-8}$ \cite{B0gg-SM}. Physics beyond 
the Standard Model could enhance this branching ratio by as much as two orders 
of magnitude\cite{B0gg-NewPhysics}.
To look for this decay, we look for events with two isolated photon candidates
with energies consistent with photons coming from the decay $ B \to \gamma \gamma$. 
As for the  $B \to K^* \gamma$ mode, 
the main backgrounds are continuum events, and we use similar requirements
to eliminate the background~\cite{PRL-B0gg}. For the purpose of determining number 
of events and efficiencies, a rectangular signal region in the $(m_{ES},\Delta E)$ 
plane is defined. Its size is determined by the $\Delta E$ and $m_{ES}$ resolution. 
The overall efficiency for $\Bogg$ events, as determined from Monte Carlo simulation,
is $(10.7 \pm 0.2)\%$. We find one event in the signal box, with an expected background
of $0.9 ^{+0.4}_{-0.3}$ events. We choose to quote a conservative upper limit on the 
branching fraction, assuming that the observed event is signal, and set the limit
$ \mathcal{B} $$(\Bogg) < 2.4\cdot 10^{-6}$ at the 90\% confidence level . This improves 
the previous limit\cite{L3-B0gg} by a factor twenty.

\end{document}
